# Towards a Multi-Objective Optimization of Subgroups for the Discovery of Materials with Exceptional Performance

Lucas Foppa
The NOMAD Laboratory at the Fritz Haber Institute of the Max-Planck-Gesellschaft and IRIS-Adlershof of the Humboldt-Universität zu Berlin
Faradayweg 4-6, 14195 Berlin, Germany
foppa@fhi-berlin.mpg.de

Matthias Scheffler
The NOMAD Laboratory at the Fritz Haber Institute of the Max-Planck-Gesellschaft and IRIS-Adlershof of the Humboldt-Universität zu Berlin
Faradayweg 4-6, 14195 Berlin, Germany
scheffler@fhi-berlin.mpg.de

**Status**

Artificial-intelligence (AI) approaches in materials science usually attempt a description of all possible scenarios with a single, global model. However, the materials that are useful for a given application, which requires a special and high performance, are often statistically exceptional. For instance, one might be interested in identifying exceedingly hard materials, or materials with band gap within a narrow range of values. Global models of materials' properties and functions are designed to perform well in average for the majority of (uninteresting) compounds. Thus, AI might well overlook the useful materials. In contrast, subgroup discovery (SGD) [1,2] identifies *local* descriptions of the materials space, accepting that a global model might be inaccurate or inappropriate to capture the useful materials subspace. Indeed, different mechanisms may govern the materials' performance across the immense materials space and SGD can focus on the mechanism(s) that result in exceptional performance.

The SGD analysis is based on a dataset $\tilde{P}$, which contains a known set of materials. $\tilde{P}$ is part of a larger space of possible materials, the full, typically infinite population $P$. For the materials in $\tilde{P}$, we know a target of interest $Y$ (metric or categorical), such as a materials' property, as well as many candidate descriptive parameters $\varphi$ possibly correlated with the underlying phenomena governing $Y$ (Fig. 1). From this dataset, SGD generates propositions $\pi$ about the descriptive parameters, e.g., inequalities constraining their values, and then identifies selectors $\sigma$, conjunctions of $\pi$, that result in SGs that maximize a quality function $Q$:

$$Q(\tilde{P}, SG) = \left(\frac{s_{SG}}{s_{\tilde{P}}}\right)^{\gamma} * \left(u(SG, \tilde{P})\right)^{1-\gamma}. \text{(Eq. 1)}$$

In Eq. 1, the ratio $s_{SG}/s_{\tilde{P}}$ is called the coverage, where $s_{SG}$ and $s_{\tilde{P}}$ are the number of data points in the SG and in $\tilde{P}$, respectively. The utility function $u(SG, \tilde{P})$ measures how exceptional the SGs are compared to $\tilde{P}$ based on the distributions of $Y$ values in the SG and in $\tilde{P}$. $Q$ establishes a tradeoff between the coverage (generality) and the utility (exceptionality), which can be tuned by a tradeoff parameter $\gamma$. Typically, the identified selectors only depend on few of the initially offered candidate descriptive parameters. The identified SG selectors (or rules) describe the local behaviour in the SG and they can be exploited for the identification of new materials in $P$.

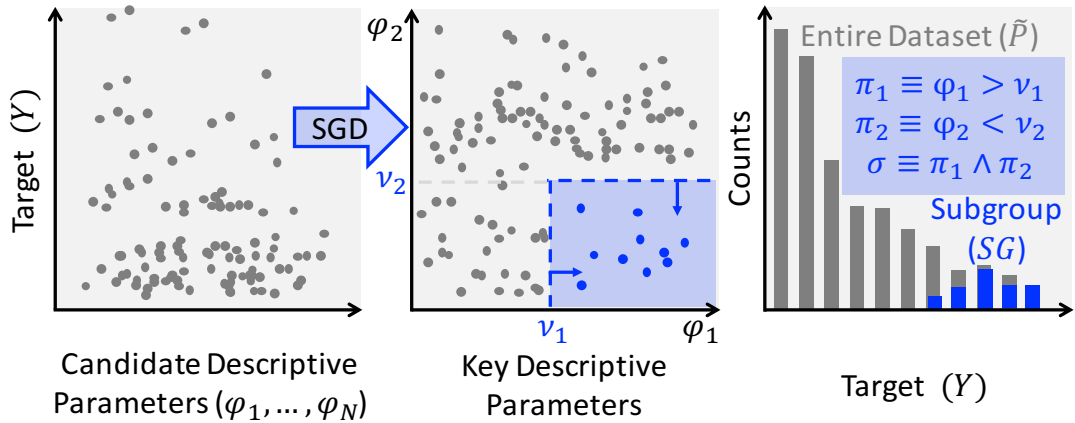

Figure 1. Subgroup discovery (SGD) identifies descriptions of exceptional subselections of the dataset. These descriptions (rules) are selectors $\sigma$ constructed as conjunctions of propositions $\pi$ about the data. The symbol ∧ denotes the "AND" operator.

**Current and Future Challenges**
The potential of SGD to uncover local patterns in materials science has been demonstrated by the identification of structure-property relationships, [3] and by the discovery of materials for heterogeneous catalysis. [4] Additionally, using (prediction) errors as target in SGD, we identified descriptions of the regions of the materials space in which (machine-learning) models have low [5] or high errors. [6] Thus, the domain of applicability (DoA) of the models could be established. Despite these encouraging results, the advancement of the SGD approach in materials science requires addressing key challenges:

- The quality function introduces one generality-exceptionality tradeoff, among a multitude of possible tradeoffs that can be relevant for a given application and that can be obtained with different $\gamma$. For instance, the required hardness of a material depends on the type of device in which it will be used and the DoA of a model depends on the accuracy that is acceptable to describe a certain property or phenomenon. However, choosing the appropriate $\gamma$ and assessing the similarity - or redundancy - among the multiple rules obtained with different tradeoffs are challenging tasks.
- Widely used utility functions assess the exceptionality of SGs by comparing the data distribution of the SG and that of $\tilde{P}$ via a single summary-statistics value. For example, the positive-mean-shift utility function for metric target favors the identification of SGs with high $Y$ values only based on the means of the two distributions. Thus, it is often assumed that the distributions are well characterized by the chosen summary-statistics value and that $\tilde{P}$ is representative of the full population $P$. However, distributions in materials science are typically non-normal and $\tilde{P}$ might not reflect the infinitely larger, unknown $P$. This calls for the consideration of utility functions that circumvent the mentioned assumptions.
- The mechanisms governing materials can be highly intricate and the relevant descriptive parameters to describe a certain materials' property are often unknown. Thus, one would like to offer many possibly relevant candidate parameters and let the SGD analysis identify the key ones. However, optimizing the quality function is a combinatorial problem with respect to the number of descriptive parameters and efficient search algorithms are therefore crucial. [7]

**Advances in Science and Technology to Meet Challenges**
In order to address some of these open questions, we approach the SGD as a multi-objective-optimization problem for the systematic identification of SG rules that correspond to a multitude of generality-exceptionality tradeoffs. Coherent collections of SG rules are obtained by considering the Pareto front of optimal SGD solutions with respect to the objectives coverage and utility function, as illustrated for the example of identification of perovskites with high bulk moduli in Fig. 2. Once the coherent collections of SG rules are identified, the overlap between SG elements can be used to assess their similarity. A high similarity between SG rules might indicate that the rules are redundant. Thus, the similarity analysis can be used to choose the SG rules that should be considered for further investigation or exploitation.

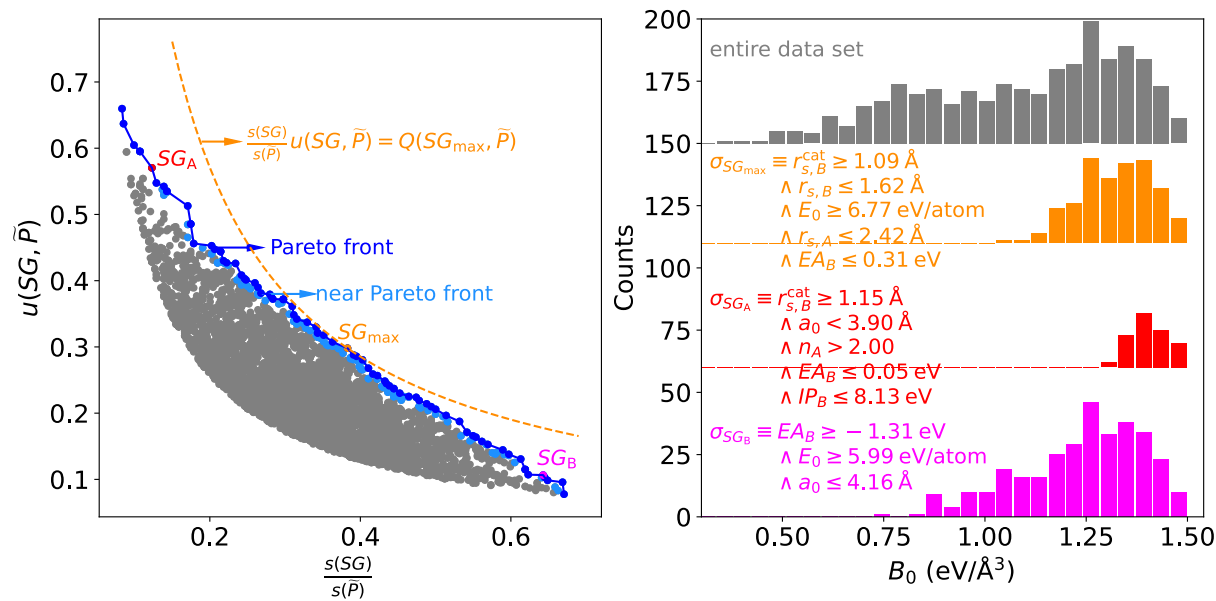

Figure 2. Left panel: A coherent collection of SG rules describing $ABO_3$ perovskites with high bulk modulus ($B_0$) is identified at the Pareto front of SGD solutions with respect to the objectives coverage and the utility function *cumulative Jensen-Shannon divergence*. Right panel: The identified rules constrain the values of the radiii of the *s* orbitals of isolated *A*, *B* and $B^{+1}$ species ($r_{s,A}$, $r_{s,B}$ and $r_{s,B}^{cat}$, respectively), the electron affinity and ionization potential of isolated *B* species ($EA_B$ and $IP_B$, respectively), the expected oxidation state of *A* ($n_A$), the equilibrium lattice constant ($a_0$), and the cohesive energy ($E_0$).

Noteworthy, the *cumulative Jensen-Shannon divergence* ($D_{JS}$) [8] between the distribution of bulk moduli in the SG and in the entire dataset is used as quality function in the example of Fig. 2. $D_{JS}$ assumes small values for similar distributions and increases as the distribution of target values in the SG is, e.g., shifted or narrower with respect to the distribution of the entire dataset. Crucially, $D_{JS}$ does not assume that one single summary-statistics value represents the distributions. Divergence-based utility functions addressing, e.g., high or low target values, will thus be an important advance. We note that the utility function might also incorporate information on multiple targets or physical constraints that are specific to the scientific question being addressed. [9] However, in order to ensure that the training data is representative of the relevant materials space one would like to cover, the iterative incorporation of new data points and training of SGD rules in an active-learning fashion might be required.

**Concluding Remarks**
SGD can accelerate the identification of exceptional materials that may be overlooked by global AI models because it focuses on local descriptions. However, further developments are required in order to translate the SGD concept to the typical scenario of materials science, where datasets might be unbalanced, or not be representative of the whole materials space and the most important descriptive parameters are unknown. The multi-objective perspective introduced in this contribution provides an efficient framework for dealing with the compromise between generality and exceptionality in SGD. The combination of this strategy with efficient algorithms for SG search and with a systematic incorporation of new data points to better cover the materials space will further advance the AI-driven discovery of materials.

**Acknowledgements**
*This work was funded by the NOMAD Center of Excellence (European Union's Horizon 2020 research and innovation program, Grant Agreement No. 951786) and by the ERC Advanced Grant TEC1p (European Research Council, Grant Agreement No. 740233).*